\documentclass[showpacs,showkeys,superscriptaddress,aps]{revtex4}
\def\be{\begin{eqnarray*}}
\def\ee{\end{eqnarray*}}
\begin{document}
\title{Intersecting D-brane states derived from the KP theory}
\author{Hironori YAMAGUCHI}
\email[email : ]{hironori@kiso.phys.metro-u.ac.jp}
\affiliation{Department of Physics, Tokyo Metropolitan University,\\
Minamiohsawa 1-1, Hachiohji, Tokyo, 192-0397 Japan}
\author{Satoru SAITO}
\email[email : ]{saito_ru@nifty.com}
\affiliation{Hakusan 4-19-10, Midoriku, Yokohama 226-0006 Japan}
\begin{abstract}

A general scheme to find tachyon boundary states is developed within the framework of the theory of KP hierarchy. The method is applied to calculate correlation function of intersecting D-branes and rederived the results of our previous works as special examples. A matrix generalization of this scheme provides a method to study dynamics of coincident multi D-branes.
\end{abstract}
\maketitle

\pagestyle{plain}
\def\be{\begin{eqnarray*}}
\def\ee{\end{eqnarray*}}
\def\ve{\vfill\eject}

\section{Introduction}

From the study of intersecting D-branes there have been known various stable configurations of D-branes. The stability of D-branes is related to intersecting angles and dimensions of the D-branes. Recently the authors of ref\cite{HN} have shown the connection between intersecting D-branes and tachyon condensation. This suggests that intersecting D-branes and recombination of D-branes have interesting feature of the string vacua.

In the dynamics of D-branes an essential part is played by the duality among string theories. In this respect one of the present authors has proved that open closed duality holds in the scheme of intersecting D-branes\cite{Y}, see also ref\cite{inter}. On the other hand we have shown in our recent paper\cite{YS} a connection between tachyon boundary states and the KP hierarchy of integrable systems. In particular we have shown that coincident D-brane states satisfy matrix generalization of Sato-Wilson type of equations\cite{Sato, SW}. This general scheme should provide us a method to analyze D-brane dynamics based on integrable systems.

The main purpose of the present paper is to rederive the intersecting D-brane states found in \cite{Y} within the frame work of integrable systems discussed in our previous paper. Our formulation of tachyon boundary states provides a quite general scheme of calculating various correlation functions. We will present, as an illustlation, a matrix generalization of our formulation to describe correlation of intersecting coincident D-branes.

A brief review of our previous works is presented in \S 2 in such a way notations are provided. In \S 3 we discuss a general framework of boundary conditions which closed strings must obey. Especially we introduce the notion of off shell tachyon boundary states and use them to derive the state of intersecting D-branes. These state thus found will be applied to calculate correlation functions in \S 4. We'll see that two results, obtained in our previous two independent works\cite{Y, SS2}, can be derived from our new formula as two special cases. The generalization to nonabelian D-branes will be discussed in the final section.

\section{A brief review of general scheme}

Let us begin our argument with preparing a general scheme of constructing tachyon boundary states which relate the D-brane dynamics to the theory of KP hierarchy. We first introduce a tachyon boundary state, which will play the central role in our discussion, and then explain its connection with the theory of integrable systems. After integrating over tachyon momenta the state will be shown to reproduce the boundary state of the de Alwis type\cite{de Alwis} as a special case and interpolates between the Dirichlet and Neumann boundary conditions in general. 

This section combines our recent works\cite{SS1, SS2, Y, YS} but arranged in such a way convenient to the discussion in the following sections. To avoid kinematical complication we do not touch on the fermionic sector in this section although it is straightforward to incorporate it into our argument. We also suppress the space-time indices in this section.

We consider the world sheet with boundaries. To study the nature of duality we introduce an index $\zeta$ and distinguish the closed string coordinate whether it is in the ordinary theory ($\zeta=1$) or in the T-dual theory ($\zeta=-1$). Hence the space-time coordinate of closed string is written as 
$$
X^\zeta(z,\bar z)=X(z)+\zeta\tilde X(\bar z).
$$
We parameterise the world sheet by $(\sigma, \tau)$, which is related to $z$ by $z=e^{\tau-i\sigma}$, and assume the boundary is at $\tau=0$. The string coordinate on the boundary is expanded according to
$$
X^\zeta(\sigma)=X_+^\zeta(\sigma)+X_-^\zeta(\sigma)
$$
\begin{equation}
X_\pm^\zeta(\sigma)
=
\left(\matrix{x_0\cr 0\cr}\right)\mp i\sum_{n=1}^\infty{1\over n}\left(\alpha_{\mp n}e^{\mp in\sigma}+\zeta\tilde\alpha_{\mp n}e^{\pm in\sigma}\right).
\label{Xpm}
\end{equation}

The bare boundary state is given by
\begin{equation}
|\rho\rangle=\prod_{n=1}^\infty\exp\left[{\rho\over n}\alpha_{-n}\tilde\alpha_{-n}\right]|0\rangle,
\label{R_b}
\end{equation}
with $\rho=1,0,-1$ corresponding to the Dirichlet, free and the Neumann boundary conditions, respectively. Owing to the introduction of $\zeta$ the boundary conditions of $X(\sigma)$ are simply given by
\begin{eqnarray}
\left(X^\zeta(\sigma)-x_0\right)|\rho\rangle
&=&
(1-\rho\zeta)\left(X^\zeta_+(\sigma)-x_0\right)|\rho\rangle,
\label{basic properties 1}\\
\partial_\tau X^\zeta(\sigma)|\rho\rangle&=&
(1+\rho\zeta)\partial_\tau X^\zeta_+(\sigma)|\rho\rangle.
\label{basic properties2}
\end{eqnarray}

In order to define a tachyon field we first generalize the single tachyon vertex operator 
$$
:e^{ikX}:=e^{ikX_+}e^{ikX_-}
$$
to
\begin{equation}
:e^{i(K,X)}:=e^{i(K,X_+)}e^{i(K,X_-)},\qquad
i(K,X)= i\int_0^{2\pi}{d\sigma\over 2\pi}\int_0^{2\pi}{d\sigma'\over 2\pi}K(\sigma)\hspace{0.5mm}\ln\left|e^{i\sigma}-e^{i\sigma'}\right|\hspace{0.5mm}{\partial X(\sigma')\over\partial\sigma'}.
\label{i(K,X)}
\end{equation}
The tachyon field is defined by
$$
\Phi(X^\zeta)=\int{\cal D}K\ \tilde\Phi_c(K) :e^{i(K,X^\zeta)}:
$$
after integrating over the momentum distribution function $K(\sigma)$. The property of the field $\Phi(X^\zeta)$ is determined by the weight function $\tilde\Phi(K)$. We assume that it has the form of
\begin{equation}
\tilde{\Phi}_{c}(K)=\prod_{n=1}^{\infty}\exp\!\bigg[-\frac{1}{2nc_{n}^{2}}p_{n}\bar{p}_{n}\bigg]. 
\label{momentum rep. of tachyon field}
\end{equation}
with $p_n$ and $\bar p_n$ being the Fourier modes of the momentum distribution function $K(\sigma)$, {\it i.e.},
$$
K(\sigma)=\sum_{n=1}^\infty\left(p_ne^{-in\sigma}+\bar p_n e^{in\sigma}\right).
$$
An arbitrary factor $c_n$ is introduced for each mode to take account the deformation from the exact Gaussian weight. We shall see in the later sections that this factor $c_n$ plays the role of parameter which interpolates different boundary conditions.

In the following discussions we employ another equivalent expression of the tachyon field $\Phi(X^\zeta)$, so that the dependence to this parameter $c_n$ becomes manifest in every step of calculations. Namely by changing the scale of each mode of momentum variables
\begin{equation}
p_{n} \rightarrow c_{n}p_{n}\hspace{1mm},\quad \bar{p}_{n} \rightarrow c_{n}\bar{p}_{n}
\end{equation}
we write $\Phi(X^\zeta)$ as
\begin{equation}
\Phi_{c}(X^\zeta)=\prod_{n=1}^{\infty}(c_{n})^{2}\int\prod_{n=1}^{\infty}i\frac{dp_{n} d\bar{p}_{n}}{4n\pi}\hspace{0.5mm}\exp\!\bigg[-\frac{1}{2n}p_{n}\bar{p}_{n}\bigg]\hspace{1mm}:e^{i(K,X^\zeta)_{c}}:\hspace{1mm}.
\label{tachyon state as our starting point}
\end{equation}
In this form the weight function is the standard Gaussian. The parameter dependence is transferred into $(K,X^\zeta)_{c}$, which is now given by 
\begin{equation}
i(K,X^\zeta)_{c}:=i\int_{0}^{2\pi}\frac{d\sigma}{2\pi}K(\sigma)\int_{0}^{2\pi}\frac{d\sigma^{\prime}}{2\pi}\frac{\partial X^\zeta(\sigma^{\prime})}{\partial\sigma^{\prime}}\Delta_{c}(\sigma,\sigma^{\prime}),
\label{phase1 of vertex op.}
\end{equation}
and
\begin{equation}
\Delta_{c}(\sigma,\sigma^{\prime}):=-\frac{1}{2}\sum_{n=1}^{\infty}\frac{c_{n}}{n}\left(e^{in(\sigma-\sigma^{\prime})}+e^{-in(\sigma-\sigma^{\prime})}\right). 
\label{deformed Green's function}
\end{equation}
is a deformed Green's function which turns back to $\ln\left|e^{i\sigma}-e^{i\sigma'}\right|$ of (\ref{i(K,X)}) at $c_{n}=1$.

The action of this field to the boundary state $|\rho\rangle$ of (\ref{R_b}) yields
\begin{equation}
\Phi_{c}(X^{\zeta})|\rho\rangle=\prod_{n=1}^{\infty}c_n^2\exp\Bigg[\rho\Big(1-(1-\zeta\rho)c_n^2\Big){1\over n}\alpha_{-n}\tilde{\alpha}_{-n}\Bigg]|0\rangle.
\label{Phi_c(X)|rho>}
\end{equation}
We notice that the Neumann and the Dirichlet boundary conditions are exchanged each other when $c_n$ is unity and $\zeta\rho=-1$. In particular the result of S.P. de Alwis\cite{de Alwis} is reproduced if we choose $\zeta\rho=-1$ and $c_{n}$ as
\begin{equation}
c_{n}=\sqrt{\frac{1}{1+\left(\frac{u}{n}\right)^{\rho}}}=\cases{\sqrt{\frac{u}{u+n}}\quad (\rho=-1) \cr \sqrt{\frac{n}{u+n}}\quad (\rho=+1)}.
\label{c_n^u}
\end{equation}

We are now going to show the relation between our formulation of the boundary states and the theory of KP hierarchy. They are related through the local property of the tachyon fields on the boundary. Therefore we study correlation functions of vertex operators instead of the tachyon field $\Phi(X^\zeta)$. In this correspondence a key role is played by the operators
$$
\phi^\zeta(z,\bar z)=:e^{{i\over \sqrt2}X^\zeta(z,\bar z)}:,\quad {\phi^*}^\zeta(z,\bar z)=:e^{{-i\over \sqrt2}X^\zeta(z,\bar z)}:
$$
which manifest the bosonization of the closed strings and satisfy 
$$
\{\phi^\zeta(z,\bar z), {\phi^*}^\zeta(z',\bar z')\}=(2\pi i)^2\delta(z-z')\delta(\bar z-\bar z').
$$
Upon exchange of $\phi^\zeta(\sigma)$ with other vertex operators we find
\begin{eqnarray}
:e^{i(K,X^\zeta)}:\phi^{\zeta'}(\sigma)&=&e^{(1-\zeta\zeta')A(\sigma)/(2\sqrt 2 i)}\phi^{\zeta'}(\sigma):e^{i(K,X^\zeta)}:\nonumber\\
A(\sigma)
&:=&
2\int_0^{2\pi}{d\sigma'\over 2\pi}K(\sigma')\Delta_c(\sigma',\sigma).
\label{A}
\end{eqnarray}
In order to relate these formulae with those of the KP theory we consider the case that the negative frequency modes of $K(\sigma)$ is given by
$$
p_n=\sum_jk_je^{in\sigma_j}\qquad n=1,2,3,...
$$
The vertex operator then turns to $:e^{i(K,X^\zeta)_c}:=\hat W^\zeta\hat V^\zeta$, where
$$
\hat V^\zeta
=
\exp\Big[{i\over\sqrt 2}\sum_jk_jX_<^\zeta(\sigma_j)\Big],\quad
\hat W^\zeta
=
\exp\Bigg[{i\over 4\pi}\oint A(\sigma)dX_>^\zeta(\sigma)\Bigg].
$$
$X_{\stackrel{<}{>}}^\zeta(\sigma)$ are the specific components of $X^\zeta$ defined by
\begin{eqnarray}
X^\zeta(\sigma)&=&X_<^\zeta(\sigma)+X_>^\zeta(\sigma),\qquad
X_{\stackrel{<}{>}}^\zeta(\sigma)
=
\left(\matrix{x_0\cr 0\cr}\right)\pm i\sum_{n=1}^\infty{1\over n}\left(\alpha_{\pm n}-\zeta\tilde\alpha_{\mp n}\right)e^{\pm in\sigma}.
\label{X<>}
\end{eqnarray}
Note that $X_>^\zeta$ ($X_<^\zeta$) component picks up only positive (negative) modes of the expansion of $K(\sigma)$. Moreover they annihilate the bare boundary states, both $|\rho\rangle$ and $\langle\rho|$, when $\rho\zeta=1$.

Using these informulae we calculate the correlation function as follows.
\begin{eqnarray}
\langle\rho'|\hat W^\zeta\hat V^\zeta\phi^{\zeta'}(\sigma)|\rho\rangle
&=&
e^{(1-\zeta\zeta')A_+(\sigma)/(2\sqrt 2 i)}e^{(1-\zeta\zeta')\xi(\sigma)/2}
\langle\rho'|\phi^{\zeta'}(\sigma)\hat W^\zeta\hat V^\zeta|\rho\rangle\\
&=&
e^{-iA_+(\sigma)/\sqrt 2}e^{\xi(\sigma)}\langle\rho'|\rho\rangle
\end{eqnarray}
when $\zeta\zeta'=-1$ and $\zeta\rho=\zeta'\rho'=1$. Here we defined
\begin{equation}
\xi(\sigma)=-\sum_jk_j\left(i\sigma_j+\sum_{n=1}^\infty{1\over n} e^{in(\sigma_j-\sigma)}\right)
\label{xi(sigma)}
\end{equation}
and $A_+(\sigma)$ is the positive frequency part of $A(\sigma)$ in (\ref{A}).

The correspondence of these formulae to those of the KP theory is accomplished if we identify the Sato-Wilson functions $\Psi_0(z)$ and $\Psi(z)$ in the KP theory\cite{Sato} with the following correlation functions\cite{YS}:
\begin{eqnarray}
\Psi_0(z)={\langle\rho'|\hat V^\zeta\phi^{\zeta'}(\sigma)|\rho\rangle\over\langle\rho'|\rho\rangle}
,\qquad
\Psi(z)={\langle\rho'|\hat W^\zeta\hat V^\zeta\phi^{\zeta'}(\sigma)|\rho\rangle\over\langle\rho'|\rho\rangle}
\label{Psi(z)=string correlator}
\end{eqnarray}
and also the variables $t_n,\ n=1,2,3,...$ of soliton equations with $k_j$'s by the Miwa transformation\cite{Miwa}
$$
t_0=i\sum_jk_j\sigma_j,\qquad t_n={1\over n}\sum_jk_je^{in\sigma_j}\quad n=1,2,3,....
$$
In terms of $t_n$'s the function $\xi(\sigma)$ of (\ref{xi(sigma)}) is written as
$$
\xi(\sigma)=-\sum_{n=0}^\infty t_ne^{-in\sigma},
$$
which is nothing but the soliton coordinate of the KP hierarchy. Accordingly the Sato-Wilson functions are given by
\begin{equation}
\Psi_0(z)=e^{\xi(\sigma)},\qquad 
\Psi(z)=W(z)\Psi_0(z)
\label{Psi}
\end{equation}
where
$$
W(z)=e^{-iA_+(\sigma)/\sqrt 2}.
$$
Since $z$ is related to $\sigma$ by $z=e^{-i\sigma}$ at the boundary, $A_+(\sigma)$, hence $W(z)$ is expanded into negative powers of $z$ as
$$
W(z)=\sum_{n=0}^\infty W_nz^{-n}.
$$
This and the special dependence of $\xi(\sigma)$ on $z$ enables us to write the second formula of (\ref{Psi}) as
$$
\Psi=W\Psi_0,\qquad W=\sum_{n=0}^\infty W_n{\partial^{-n}\over\partial t_1^{-n}}.
$$

In the theory of KP hierarchy the Sato-Wilson function $\Psi(z)$ appears as a solution to the eigenvalue problems for given Lax pairs $L$ and $B_n$
\begin{equation}
L\Psi=z\Psi,\quad {\partial\Psi\over\partial t_n}=B_n\Psi,\quad n=0,1,2,....
\label{Lax eq}
\end{equation}
The pseudo differential operator $W$, which is also called Sato-Wilson operator, is then related with $L$ through
$$
L=W{\partial\over \partial t_1}W^{-1}.
$$

One of advantages of this formulation of the KP hierarchy is that it can be generalized into a matrix form straightforwardly\cite{Sato}. Translated into the terminology of the closed strings it amounts to generalize $k_j$'s to diagnal matrices and $A_+(\sigma)$ to a general matrix. In our previous paper\cite{YS} we suggested to interprete this fact as a result of the existence of coincident multi D-branes.

\section{Solutions to the boundary conditions}

From the discussions in the previous section we see that the important ingredient of our study is the tachyon boudary state $:e^{i(K,X^\zeta)}:|\rho\rangle$. It constitutes the source of tachyon condensation on the boundary. At the same time it relates the theory of closed strings directly to the theory of integrable systems. Let us denote this state simply as
\begin{equation}
|A,\rho\rangle=\ :\exp\Bigg[{i\over 4\pi}\oint A^\mu(\sigma) dX_\mu^\zeta(\sigma)\Bigg]:|\rho\rangle.
\label{|A>}
\end{equation}
Here $A^\mu(\sigma)$ is the one related to the momentum distribution $K^\mu(\sigma)$ by (\ref{A}). We recovered the spacetime index $\mu$, since we are interested in the spacetime structure of D-branes in this section. Note that the values of $\rho$ in the state $|\rho\rangle$ can be different depending on the direction of spacetime. It must be also mentioned that the index $\zeta$ of $X^\zeta$ in (\ref{|A>}) is suppressed from $|A,\rho\rangle$ because it is fixed, due to the relation of (\ref{basic properties 1}), by $\zeta\rho=-1$ when $\rho$ is given. 

An important feature of this state is that this is an eigenstate of the spacetime coordinate $X_\mu^{\zeta_o}(\sigma)$ on the boundary when $\zeta_o\rho_\mu=1$. In fact we have
\begin{eqnarray}
X_\mu^{\zeta_o}(\sigma)|A,\rho\rangle
&=&
A^\mu(\sigma)|A,\rho\rangle\qquad {\rm if}\quad \zeta_o\rho_\mu=1.
\label{eigeneq of X}\\
\partial_\tau X_\mu^{\zeta_o}(\sigma)|A,\rho\rangle
&=&
\partial_\tau A^\mu(\sigma)|A,\rho\rangle\qquad {\rm if}\quad \zeta_o\rho_\mu=-1.
\label{eigeneq of dX}
\end{eqnarray}
We must emphasize here that (\ref{eigeneq of X}) is true only when $\zeta_o\zeta=-1$, while (\ref{eigeneq of dX}) holds when $\zeta_o\zeta=1$. In other words the boundary state $|A,\rho\rangle$ is an eigenstate of $\partial_\tau X^{\zeta}$ and also of $X^{\zeta_o}=X^{-\zeta}$, which is dual to $X^\zeta$ in the state $|A,\rho\rangle$. 

We can define fermionic partner of the tachyon boundary state quite parallel to the bosonic case. Let us consider the NS coordinate $\psi_\mu^\zeta(\sigma)$ defined by
\begin{eqnarray}
\psi_\mu^\zeta(\sigma)&=&e^{i\sigma/2}\sum_{r={1\over 2}}^\infty\left(\psi_r^\mu e^{ir\sigma}+\psi_{-r}^\mu e^{-ir\sigma}+i\zeta\tilde\psi_r^\mu e^{-ir\sigma}+i\zeta\tilde\psi_{-r}^\mu e^{ir\sigma}\right)
\end{eqnarray}
and a fermionic boundary state ($\eta$ denotes the spin structure)
$$
|\Theta,\rho,\eta\rangle
=
\exp\Bigg[-{1\over 4\pi}\int_0^{2\pi}\Theta^\mu(\sigma)\psi^\zeta_\mu(\sigma)e^{-i\sigma/2}d\sigma\Bigg]|\rho,\eta\rangle\quad {\rm with}\quad \eta\rho\zeta=1.
$$
Here $\Theta^\mu(\sigma)$ is the fermionic counterpart of $A^\mu(\sigma)$ expanded as 
\begin{equation}
\Theta^\mu(\sigma)=\int_0^{2\pi}{d\sigma'\over 2\pi}{\cal K}^\mu(\sigma')\sum_{r={1\over 2}}^\infty c_r(e^{ir(\sigma-\sigma')}+e^{ir(\sigma-\sigma')})
\end{equation}
Then we can convince ourselves that the following relations hold:
\begin{eqnarray}
\psi_\mu^{\zeta_o}(\sigma)|\Theta,\rho,\eta\rangle
&=&
\Theta^\mu(\sigma)|\Theta,\rho,\eta\rangle\qquad {\rm if}\quad \eta\rho\zeta_o=-1
\label{psi|Theta>=Theta|Theta}\\
\partial_\tau\psi_\mu^{\zeta_o}(\sigma)|\Theta,\rho,\eta\rangle
&=&
\partial_\tau\Theta^\mu(\sigma)|\Theta,\rho,\eta\rangle\qquad {\rm if}\quad \eta\rho\zeta_o=1.
\label{dpsi|Theta>=dTheta|Theta}
\end{eqnarray}

Now we would like to know the state of intersecting D-branes. For this purpose let us first derive a state of D-brane whose normal to a flat surface is rotated in the $(i, i+p)$ plane by the angle $\phi$ from the $i+p$ axis. The state must satisfy Dirichlet condition along the normal while it must satisfy Neumann condition along the surface. Hence we impose the conditions
\begin{eqnarray}
\left(\sin\phi X_i^{\zeta_o}(\sigma)+\cos\phi X_{{i+p}}^{\zeta_o}(\sigma)\right)|A,\rho\rangle&=&0,
\label{sX+CX}\\
\left(\cos\phi\ \partial_\tau X_i^{\zeta_o}(\sigma)-\sin\phi\ \partial_\tau X_{{i+p}}^{\zeta_o}(\sigma)\right)|A,\rho\rangle&=&0,
\label{cX-sX}\\
\left(\sin\phi\ \psi_i^{\zeta_o}(\sigma)+\cos\phi\ \psi_{{i+p}}^{\zeta_o}(\sigma)\right)|\Theta,\rho,\eta\rangle&=&0,
\label{spsi+cpsi}\\
\left(\cos\phi\ \partial_\tau \psi_i^{\zeta_o}(\sigma)-\sin\phi\ \partial_\tau \psi_{{i+p}}^{\zeta_o}(\sigma)\right)|\Theta,\rho,\eta\rangle&=&0.
\label{cpsi-spsi}
\end{eqnarray}
By (\ref{eigeneq of X}) and (\ref{psi|Theta>=Theta|Theta}) we can find solutions to (\ref{sX+CX}) as well as (\ref{spsi+cpsi}) easily. They are satisfied if 
\be
\sin\phi\ A^i(\sigma)+\cos\phi\ A^{i+p}(\sigma)&=&0\\
\sin\phi\ \Theta^i(\sigma)+\cos\phi\ \Theta^{i+p}(\sigma)&=&0
\ee
hold. Similarly (\ref{cX-sX}) and (\ref{cpsi-spsi}) are satisfied if
\be
\cos\phi\ \partial_\tau A^i(\sigma)-\sin\phi\ \partial_\tau A^{i+p}(\sigma)&=&0\\
\cos\phi\ \partial_\tau \Theta^i(\sigma)-\sin\phi\ \partial_\tau \Theta^{i+p}(\sigma)&=&0
\ee
hold. It is certainly impossible for $|A,\rho\rangle$ and $|\Theta,\rho,\eta\rangle$ to satisfy these two conditions simultaneously, since they are `plane wave solutions' propagating to the directions orthogonal to each other. 

In order to obtain a simultaneous solution of both (\ref{sX+CX}) and (\ref{cX-sX}), we apply the standard technique for solving a wave propagation in 2 dimensions under two independent boundary conditions. Namely we superpose the plane waves satisfying (\ref{sX+CX}) and determine the weight of each mode such that another condition (\ref{cX-sX}) is satisfied. In doing it we recall the fact that the state $|A,\rho\rangle$ contains parameters $c_n$ which interpolates two different boundary conditions. In our present case we have right to choose them for each direction of spacetime independently.

Using the notations
$$
A^i X_i+A^{i+p} X_{i+p}=A^i X_\phi,\qquad X_\phi:=X_i-{\sin\phi\over\cos\phi} X_{i+p}
$$
the boundary state of the closed string satisfying (\ref{sX+CX}) is given by
$$
|A_\phi,\rho\rangle:=\exp\Bigg[{i\over 2\pi}\oint A^i(\sigma)dX^\zeta_{+\phi}(\sigma)\Bigg]|\rho\rangle
$$
where the Wilson loop integral of the state is
$$
{i\over 2\pi}\oint A^i(\sigma)dX^\zeta_{+\phi}(\sigma)
=
i\sum_{n=1}^\infty{c^i_n\over n}\left(\bar p_n^i\Bigg(\alpha_{-n}^i-{\sin\phi\over\cos\phi} \alpha_{-n}^{i+p}\Bigg)
-\zeta p_n^i\Bigg(\tilde\alpha_{-n}^i-{\sin\phi\over\cos\phi} \tilde\alpha_{-n}^{i+p}\Bigg)\right).
$$
Integrating over $p_n$ and $\bar p_n$ with the standard Gaussian weight we find
\begin{eqnarray}
\Phi_c(X^\zeta_\phi)|\rho\rangle
&:=&
\int\prod_{n=1}^\infty{dp^i_n d\bar p_n^i\over 4n\pi}\exp\Bigg[-{1\over 2n}p_n^i\bar p_n^i\Bigg]\exp\Bigg[{i\over 2\pi}\oint A^i(\sigma)dX^\zeta_{+\phi}(\sigma)\Bigg]|\rho\rangle
\nonumber\\
&=&
\prod_{n=1}^\infty\exp\Bigg[{2\zeta(c_n^i)^2\over n}\Bigg(\alpha_{-n}^i-{\sin\phi\over\cos\phi} \alpha_{-n}^{i+p}\Bigg)\Bigg(\tilde\alpha_{-n}^i-{\sin\phi\over\cos\phi} \tilde\alpha_{-n}^{i+p}\Bigg)\Bigg]|\rho\rangle
\label{Phi|rho>}
\end{eqnarray}
with $\zeta=-\rho$. When $c_n^i={\cos\phi}$ for all $n$ it turns to
\begin{equation}
\prod_{n=1}^\infty\exp\Bigg[{\rho\over n}\Bigg(\cos 2\phi\Big( \alpha_{-n}^{i+p}\tilde\alpha_{-n}^{i+p}-\alpha_{-n}^i\tilde\alpha_{-n}^i\Big)
+\sin 2\phi\Big(\alpha_{-n}^{i+p}\tilde\alpha_{-n}^i+\alpha_{-n}^i\tilde\alpha_{-n}^{i+p}\Big)\Bigg)\Bigg]|0\rangle.
\label{bose corr}
\end{equation}

Applying almost the same argument to the fermionic case, we obtain
\begin{equation}
\Phi_c(\psi_\phi^\zeta)|\rho,\eta\rangle
=
\prod_{r={1\over 2}}^\infty\exp\Big[-2i\zeta (c^i_r)^2\Bigg(\psi_{-r}^i-{\sin\phi\over\cos\phi} \psi_{-r}^{i+p}\Bigg)\Bigg(\tilde\psi_{-r}^i-{\sin\phi\over\cos\phi} \tilde\psi_{-r}^{i+p}\Bigg)\Big]|\rho,\eta\rangle,
\end{equation}
with $\zeta=\eta\rho$, which turns to
\begin{equation}
\prod_{r={1\over 2}}^\infty\exp\Bigg[i\eta\rho\Bigg(\cos 2\phi\Big( \psi_{-r}^{i+p}\tilde\psi_{-r}^{i+p}-\psi_{-r}^i\tilde\psi_{-r}^i\Big)
+\sin 2\phi\Big(\psi_{-r}^{i+p}\tilde\psi_{-r}^i+\psi_{-r}^i\tilde\psi_{-r}^{i+p}\Big)\Bigg)\Bigg]|0\rangle,
\label{fermion corr}
\end{equation}
when $c^i_r=\cos\phi$.
These formula (\ref{bose corr}) and (\ref{fermion corr}) are the result we derive from the tachyon boundary state and satisfy all conditions from (\ref{sX+CX}) to (\ref{cpsi-spsi}) simultaneously. They are also exactly the same with what we found in \cite{Y} through trial and error.

Instead of (\ref{sX+CX}) and (\ref{spsi+cpsi}) we could start our derivation of the state from the condition (\ref{cX-sX}) and (\ref{cpsi-spsi}) first. If we did so, we obtained the following state
\begin{equation}
\Phi'_c(X_\phi^\zeta)|\rho\rangle
=
\prod_{n=1}^\infty\exp\Bigg[{2\zeta(c_n^i)^2\over n}\Bigg(\alpha_{-n}^i+{\cos\phi\over\sin\phi} \alpha_{-n}^{i+p}\Bigg)\Bigg(\tilde\alpha_{-n}^i+{\cos\phi\over\sin\phi} \tilde\alpha_{-n}^{i+p}\Bigg)\Bigg]|\rho\rangle
\label{start from dX bc}
\end{equation}
when $\zeta\rho=-1$, and 
\begin{equation}
\Phi'_c(\psi^\zeta_\phi)|\rho,\eta\rangle
=
\prod_{r={1\over 2}}^\infty\exp\Big[-2i\zeta (c^i_r)^2\Bigg(\psi_{-r}^i+{\cos\phi\over\sin\phi} \psi_{-r}^{i+p}\Bigg)\Bigg(\tilde\psi_r^i+{\cos\phi\over\sin\phi} \tilde\psi_{-r}^{i+p}\Bigg)\Big]|\rho,\eta\rangle,
\label{fermion corr2}
\end{equation}
when $\rho\zeta\eta=1$. If set we$c^i_n=c^i_r=\sin\phi$ we obtain exactly (\ref{bose corr}) and (\ref{fermion corr}) but only $\rho$ is replaced by $-\rho$. This means that we obtained two different solutions to the boundary state problem (\ref{sX+CX}) through (\ref{cpsi-spsi}) with different values of $\rho$, corresponding either Dirichlet or Neumann conditions. This difference, however, is irrelevant since it changes only over all factor in the conditions.

\section{Correlation functions of intersecting D-branes}

Having found solutions to the boundary conditions we can calculate correlation functions of intersecting D-branes. Suppose there are two D-branes intersecting each other. We set the normal of one of the branes rotated by angle $\phi$ from the $i+p$ axis in the $(i,i+p)$ plane as before and another by angle $\phi'$. The quantity which we are going to calculate is 
$$
\langle\rho'|\Phi_{c'}^\dagger(X^{\zeta'}_{\phi'})e^{-l(L_0+\tilde L_0)}\Phi_{c}(X^{\zeta}_{\phi})|\rho\rangle.
$$
Instead of substituting $\Phi(X)$'s of (\ref{Phi|rho>}) into the above formula, however, it is more interesting to calculate the correlation functions of local quantities $e^{i\oint AdX}$ first. Since $A^\mu(\sigma)$ is the eigenvalue of the spacetime coordinate of the boundary we can expect to extract some local informations about D-branes. 

The general formulae given in Appendix of \cite{SS2} are quite useful to simplify the calculation of correlators of this type. The formulae we use are
\be
&&\langle 0|\exp\Bigg[{f\over n}\alpha_n\tilde\alpha_n\Bigg]\exp(\alpha_na+\tilde\alpha_n\bar a)\exp({b\alpha_{-n}+\bar b\tilde\alpha_{-n}})\exp\Bigg[{g\over n}\alpha_{-n}\tilde\alpha_{-n}\Bigg]|0\rangle\\
&&=
{1\over 1-fg}\exp\Bigg[{n\over 1-fg}(ab+\bar a\bar b+ga\bar a+fb\bar b)\Bigg]
\ee
for the bosonic sector and
\be
&&\langle 0|\exp({f\psi_r\tilde\psi_r})\exp({\psi_ra+\tilde\psi_r\bar a})\exp({b\psi_{-r}+\bar b\tilde\psi_{-r}})\exp({g\psi_{-r}\tilde\psi_{-r}})|0\rangle\\
&&=
(1-fg)\exp\Bigg[{1\over 1-fg}(ab+\bar a\bar b+ga\bar a+fb\bar b)\Bigg]
\ee
for the fermionic sector.

Using these formulae we find the propagator of the closed string between two boundary states as
\begin{eqnarray}
&&\langle A'_{\phi'},\rho'|e^{-l(L_0+\tilde L_0)}|A_\phi,\rho\rangle\nonumber\\
&&=
\prod_{n=1}^\infty{1\over 1-\rho\rho'e^{-2nl}}
\exp\Bigg[\int_0^{2\pi}{d\sigma\over 2\pi}\int_0^{2\pi}{d\sigma'\over 2\pi}\Bigg(-{\cos(\phi-\phi')\over\cos\phi\cos\phi'}{A'}(\sigma)G_1(\sigma,\sigma')A(\sigma')
\nonumber\\
&&\qquad
+
{1\over \cos^2\phi'}{A'}(\sigma)G_2(\sigma,\sigma'){A'}(\sigma')+{1\over \cos^2\phi}A(\sigma)G_2(\sigma,\sigma')A(\sigma')
\Bigg)\Bigg]
\label{correlator}
\end{eqnarray}
where
$$
G_1=\sum_{n=1}^\infty n e^{nl}{e^{in(\sigma-\sigma')}+\rho\rho'e^{-in(\sigma-\sigma')}\over 1-\rho\rho'e^{2nl}},\qquad
G_2=\sum_{n=1}^\infty n {e^{in(\sigma-\sigma')}\over 1-\rho\rho'e^{2nl}}.
$$
We supressed the spacetime index in this expression since all of the spacetime component of the vectors are only one of the $i$th direction. Because $A(\sigma)$ is the eigenvalue of the spacetime coordinate $X(\sigma)$, we can extract some local informations of the D-branes from this expression. For example the interaction term $A'A$ disappears when the angle between two D-branes $\phi-\phi'$ is $\pi/2$. Writing them in components we have
\be
&&
\langle A'_{\phi'},\rho'|e^{-l(L_0+\tilde L_0)}|A_\phi,\rho\rangle=
\prod_{n=1}^\infty{1\over 1-\rho\rho'e^{-2nl}}\\
&&
\times
\exp\left[{\displaystyle{
{c_n^2 \over\cos^2\phi}}p_n\bar p_n+\displaystyle{{{c'}_n^2 \over\cos^2\phi'}}p'_n\bar p'_n-\displaystyle{{c_n{c'}_ne^{nl}\cos(\phi-\phi')\over\cos\phi\cos\phi'}}(\bar p'_np_n+\rho\rho' p'_n\bar p_n)
\over n(1-\rho\rho' e^{2nl})}\right]
\ee

The fermionic part can be also calculated almost parallel and obtain
\be
&&\langle \Theta'_{\phi'},\rho',\eta'|e^{-l(L_0+\tilde L_0)}|\Theta_\phi,\rho,\eta\rangle\\
&=&
\prod_{r={1\over 2}}^\infty (1-\eta\eta'\rho\rho'e^{-2rl})
\exp\Bigg[\int_0^{2\pi}{d\sigma\over 2\pi}\int_0^{2\pi}{d\sigma'\over 2\pi}\Bigg(-{\cos(\phi-\phi')\over\cos\phi\cos\phi'}{\Theta'}(\sigma){\cal G}_1(\sigma,\sigma')\Theta(\sigma')\\
&&\qquad
+
{1\over \cos^2\phi'}{\Theta'}(\sigma){\cal G}_2(\sigma,\sigma'){\Theta'}(\sigma')+{1\over \cos^2\phi}\Theta(\sigma){\cal G}_2(\sigma,\sigma')\Theta(\sigma')
\Bigg)\Bigg]\\
&=&
\prod_{r={1\over 2}}^\infty (1-\eta\eta'\rho\rho'e^{-2rl})\\
&&
\times
\exp\left[{\displaystyle{
{c_r^2 \over\cos^2\phi}}\theta_r\bar \theta_r+\displaystyle{{{c'}_r^2 \over\cos^2\phi'}}\theta'_r\bar \theta'_r-\displaystyle{{c_r{c'}_re^{rl}\cos(\phi-\phi')\over\cos\phi\cos\phi'}}(\bar \theta'_r\theta_r+\eta\eta'\rho\rho' \theta'_r\bar \theta_r)
\over (1-\eta\eta'\rho\rho' e^{2rl})}\right]
\ee
where
$$
{\cal G}_1=\sum_{r={1\over 2}}^\infty e^{rl}{e^{ir(\sigma-\sigma')}+\eta\eta'\rho\rho'e^{-ir(\sigma-\sigma')}\over 1-\eta\eta'\rho\rho'e^{2rl}},\qquad
{\cal G}_2=\sum_{r={1\over2}}^\infty {e^{ir(\sigma-\sigma')}\over 1-\eta\eta'\rho\rho'e^{2rl}}.
$$

The full contribution of the propagation of closed strings will be obtained by integrating over the momentum distribution with the Gaussian weights for both bosonic and fermionic sectors separately. Combining them together the correlator of the tachyon fields becomes
\begin{eqnarray}
&&
\langle\rho',\eta'|\Phi_{c'}^\dagger(X_{\phi'},\psi_{\phi'})e^{-l(L_0+\tilde L_0)}\Phi_c(X_\phi,\psi_{\phi})|\rho,\eta\rangle\nonumber\\
&&=
\prod_{\stackrel{n=1}{r={1\over 2}}}^\infty
{c^2_n{c'}^2_n\over c_r^2{c'}_r^2}
{\left(1-\displaystyle{{2c_r^2\over\cos^2\phi}}\right)\left(1-\displaystyle{{2{c'}_r^2\over\cos^2\phi'}}\right)-\eta\eta'\rho\rho'e^{2rl}\left(1-
\displaystyle{{4c_r^2{c'}_r^2\over\cos^2\phi\cos^2\phi'}{\sin^2(\phi-\phi')\over 1-\eta\eta'\rho\rho'e^{2rl}}}\right)
\over
\left(1-\displaystyle{{2c_n^2\over\cos^2\phi}}\right)\left(1-\displaystyle{{2{c'}_n^2\over\cos^2\phi'}}\right)-\rho\rho'e^{2nl}\left(1-
\displaystyle{{4c_n^2{c'}_n^2\over\cos^2\phi\cos^2\phi'}{\sin^2(\phi-\phi')\over 1-\rho\rho'e^{2nl}}}\right)}.
\end{eqnarray}
This reproduces the result of \cite{SS2}, {\it i.e.,} 
$$
\prod_{\stackrel{n=1}{r={1\over 2}}}^\infty
{c^2_n{c'}^2_n\over c_r^2{c'}_r^2}
{(1-2c_r^2)(1-2{c'}_r)-\eta\eta'\rho\rho'e^{2rl}\over
(1-2c_n^2)(1-2{c'}_n)-\rho\rho'e^{2nl}}
$$
in the case of $\phi=\phi'=0$, hence with no intersection. When $c_n=c_r=\cos\phi$ and $c'_n=c'_r=\cos\phi'$, it is
$$
\prod_{\stackrel{n=1}{r={1\over 2}}}^\infty
{1-\rho\rho'e^{2nl}\over 1-\eta\eta'\rho\rho'e^{2rl}}\ 
{\Big(1-\eta\eta'\rho\rho'e^{2rl+2i(\phi-\phi')}\Big)\Big(1-\eta\eta'\rho\rho'e^{2rl-2i(\phi-\phi')}\Big)\over
\Big(1-\rho\rho'e^{2nl+2i(\phi-\phi')}\Big)\Big(1-\rho\rho'e^{2nl-2i(\phi-\phi')}\Big)}.
$$
This is the on-shell result derived in our paper \cite{Y}. There it was shown that this agrees with the one calculated from the open string channel after the duality transformation.

\section{Matrix generalization}

As we mentioned in \S 2 our scheme of formulation of tachyon boundary states admits a generalization to the matrix form within the framework of integrable systems. The purpose of this section is to explain how it should be done. It was shown in our previous paper that the Sato-Wilson formalism of the KP hierarchy can be rephrased in terms of the language of string theory. This correspondence enabled us to find a fermionic partner of the KP theory. Moreover, since the generalization of the Sato-Wilson formalism was well known, we introduced quite naturally a matrix generalization of the tachyon boundary states.

The correspondence of the Sato-Wilson functions with the string correlation functions has been presented in (\ref{Psi(z)=string correlator}). The generalization of them to the matrix form is then given by replacing $\hat V$ and $\hat W$ by
\begin{eqnarray}
{\hat{\bm{V}}}^\zeta&=&\exp\left[{i\sqrt2}\sum_j\left({\bm{k}}_jX^\zeta_<(\sigma_j)-i{\bm{\kappa}}_j\psi^\zeta_<(\sigma_j)\right)\right],
\label{matrix hat V}\\
{\hat{\bm{W}}}^\zeta
&=&\exp\left[{i\over 4\pi}\oint\bm{A}(\sigma)dX^\zeta_>(\sigma)-{1\over 4\pi}\int_0^{2\pi} \bm{\Theta}(\sigma)\psi^\zeta_>(\sigma)e^{-i\sigma/2}d\sigma\right]
\label{matrix hat}
\end{eqnarray}
Here we denote by boldface the matrices. We also incorporated the fermionic partners into the formulae. To obtain the Sato-Wilson matrix amplitudes, $\bm{k}_j$'s and $\bm{\kappa}_j$'s are set to be diagonal matrices. Recall that the soliton coordinate $\xi(\sigma)$ in the KP theory appears as an eigenvalue of the quantity $\sum_j{\bm{k}}_jX^\zeta_<(\sigma_j)$ in $\hat{\bm{V}}$. On the other hand $\bm{A}$ and $\bm{\Theta}$ are matrices in general and behave as nonabelian gauge fields.

Once we have established the correspondence it is natural to consider more general amplitude
$$
\bm{\Psi}(z)=\langle \bm{A},\bm{\Theta},\eta',\rho'|\phi^{\zeta}(z)|\eta,\rho\rangle
$$
with the state $|\bm{A},\bm{\Theta},\eta,\rho\rangle$ being defined by
$$
|\bm{A},\bm{\Theta},\eta,\rho\rangle=
\exp\left[{i\over 4\pi}\oint\bm{A}(\sigma)dX^\zeta(\sigma)-{1\over 4\pi}\int_0^{2\pi} \bm{\Theta}(\sigma)\psi^\zeta(\sigma)e^{-i\sigma/2}d\sigma\right]|\eta,\rho\rangle.
$$
The Sato-Wilson amplitude is obtained by separating $X(\sigma)$ into the positive and negative frequency parts, $X_<$ and $X_>$, according to (\ref{X<>}), and if the negative frequency part of the matrix $\bm{A}$ is restricted to a diagonal one. They will be related by a gauge transformation. 

The vertex operator $\phi^\zeta(z)$ in $\bm{\Psi}(z)$ plays the role of test charge, by which we can study the nature of the state $|\bm{A},\bm{\Theta},\eta,\rho\rangle$. The Sato-Wilson equation of the KP theory determines the properties which they must obey. Usig solutions to the equation we can evaluate various physical quantities, such as correlation functions. Our tachyon boundary states are exactly such states.

In order to proceed further we must know how to calculate the boundary states and correlation functions when the gauge fields are nonabelian. The matrix generalization of the theory of KP hierarchy was studied in general linear matrices. From the point of view of D-brane dynamics we are interested in nonabelian gauge theory which arises from coincident D-branes. Therefore we suppose $\bm{A}$ and $\bm{\Theta}$ are elements of $\bm{u}(N)$ and write them as
$$
\bm{A}(\sigma)=\sum_{a}\bm{\lambda}^aA_a(\sigma),\quad
\bm{\Theta}(\sigma)=\sum_{a}\bm{\lambda}^a\Theta_a(\sigma)
$$
with $\bm{\lambda}^a$ being the generators of $U(N)$. Here we supressed the spacetime indices again and added the indices of $U(N)$.

The boundary state in which tachyon is condensed should be given by
$$
\bm{\Phi}(X^\zeta,\psi^\zeta)={\rm Ptr}\ \exp\Bigg[i{\bm{\lambda}^a\over 4\pi}\Bigg(\oint A_a(\sigma)dX^\zeta(\sigma)+i\int_0^{2\pi} \Theta_a(\sigma)\psi^\zeta(\sigma)e^{-i\sigma/2}d\sigma\Bigg)\Bigg]|\eta,\rho\rangle.
$$
Where the symbol Ptr means the standard path ordered trace including the path integration over $A_a(\sigma)$ and $\Theta_a(\sigma)$ with proper Gaussian weights.

Finally we can apply our formulation to describe intersecting multi D-branes. The correlation function of intersecting D-branes was given by (\ref{correlator}). The generalization to nonabelian case is straightforward. The object now describes $N$ coincident D-branes intersecting each other, and is given by
\begin{eqnarray}
&&\langle \bm{A}'_{\phi'},\rho'|e^{-l(L_0+\tilde L_0)}|\bm{A}_\phi,\rho\rangle\nonumber\\
&&=\prod_{n=1}^\infty{1\over 1-\rho\rho'e^{-2nl}}\exp\Bigg[\bm{\lambda}^a\bm{\lambda}^b\int_0^{2\pi}{d\sigma\over 2\pi}\int_0^{2\pi}{d\sigma'\over 2\pi}\Bigg(-{\cos(\phi-\phi')\over\cos\phi\cos\phi'}{A'_a}(\sigma)G_1(\sigma,\sigma')A_b(\sigma')
\nonumber\\
&&\qquad\qquad
+
{1\over \cos^2\phi'}{A'_a}(\sigma)G_2(\sigma,\sigma'){A'_b}(\sigma')+{1\over \cos^2\phi}{A_a}(\sigma)G_2(\sigma,\sigma')A_b(\sigma')
\Bigg)\Bigg]
\label{matrix correlator}
\end{eqnarray}
The second and the third terms in the exponent represent the interaction of coincident D-branes in the same group while the first term represents the exchange of information between intersecting D-branes. The strength of the interaction depends on the intersection angle $\phi-\phi'$ and disappears when the D-branes are orthogonal with each other.



\begin{thebibliography}{99}

\bibitem{HN}
K.Hashimoto and S.Nagaoka, JHEP {\bf 0306}, 034 (2003).
\bibitem{Y}
H.Yamaguci, Open-closed duality of intersecting branes, Mod. Phys. Lett. {\bf A19} (2004) 2353-2363.

\bibitem{YS}
H.Yamaguchi and S. Saito, hep-th/0412056.
\bibitem{Sato}
M.Sato, Publ. RIMS {\bf 433} (1981) 30-46.
\bibitem{SW}
G.Wilson, On two construtions of conservation laws for Lax equation. Quart. J. Math. Oxford {\bf 32} (1981) 491-512.
\bibitem{SS2}
R.Sato and S.Saito, D-brane correlators as solutions of Hirota-Miwa equation, JHEP {\bf 0411} (2004) 047, hep-th/0408149.

\bibitem{de Alwis}
S.P.de Alwis, Phys.Lett. {\bf B505} (2001) 215-221.
\bibitem{SS1}
S.Saito and R.Sato, Soliton equations solved by the boundary CFT, JHEP {\bf 0311} (2003) 008, hep-th/0307159.

\bibitem{Miwa}
T.~Miwa, On Hirota's difference equations, Proc. Japan Acad. A {\bf 58} (1982) 9-12.

\bibitem{inter}
M. Berkooz, M.R. Douglas, R.G. Leigh, Branes Intersecting at Angles
, Nucl.Phys.B480:265-278,1996, hep-th/9606139;\\ 
H. Arfaei, M.M. Sheikh Jabbari, Different D-brane Interactions, Phys.Lett. B394 (1997) 288-296, hep-th/9608167;\\
M.M. Sheikh-Jabbari, Classification of Different Branes at Angles, Phys.Lett. B420 (1998) 279-284, hep-th/9710121;\\
T. Kitao, N. Ohta, Jian-Ge Zhou, Fermionic Zero Mode and String Creation between D4-Branes at Angles, Phys.Lett. B428 (1998) 68-74, hep-th/9801135
\end{thebibliography}
\end{document}